# Quickest Attack Detection in Smart Grid Based on Sequential Monte Carlo Filtering

Leian Chen and Xiaodong Wang

**Abstract**

Quick and accurate detection of cyber attack is key to the normal operation of the smart grid system. In this paper, a joint state estimation and sequential attack detection method for a given bus with grid frequency drift is proposed that utilizes the commonly monitored output voltage. In particular, based on a non-linear state-space model derived from the three-phase sinusoidal voltage equations, we employ the sequential Monte Carlo (SMC) filtering to estimate the system state. The output of the SMC filter is fed into a CUSUM test to detect the attack in a fastest way. Moreover, an adaptive sampling strategy is proposed to reduce the rate of taking measurements and communicating with the controller. Extensive simulation results demonstrate that the proposed method achieves high adaptivity and efficient detection of various types of attacks in power systems.

## I. Introduction

The smart grid transforms the legacy power grid that provides a one-way centrally generated power flow to end users into a more distributed and dynamic system of two-way flow of power and information [1]. In a large-scale power system, the increased connectivity and communication load lead to severe security challenges arising from physical faults and malicious attacks. The occurrence of cyber attacks can result in energy loss and safety concerns [2] [3]. Therefore, there is an urgent need for effective and economical mechanisms to detect structured attacks and to safeguard the smart grid.

### A. Related Works

Typical attacks interfere with the data communication in a smart grid system by jamming the channel, injecting malicious data, etc. To protect the smart grid from external attacks, existing approaches include early anomaly detection methods that identify faulty events in advance [4] [5], encryption and authentication schemes adopted from Internet communications [6]–[8], and control-theoretic methods based on the state estimation [9]–[14]. Here, we focus on the control-theoretic methods, considering the limitations of the first two approaches, e.g., heavy dependence on numerous feedback reports and continuous calibration of the whole system, computational and communication requirements beyond the ability of local devices.

A general control-theoretic approach is to jointly deploy an estimator and a detector. The estimator utilizes the actual measurement readings to "predict" the system state, and the detector triggers the alarm when the estimated states deviate from the actual ones. In [9], the authors propose to detect data injection attacks by tracking the dynamics of measurement variations represented by the Kullback-Leibler distances between adjacent steps. [10] adopts the Kalman filter to estimate the state variables when the grid frequency is constant, and proposes a Euclidean detector to cope with sophisticated injection attack. [11] focuses on the state estimation and detection of unobservable sparse data injection attacks. Machine learning methods have also been proposed to decide whether a given measurement is normal or attacked, while its accuracy is heavily dependent on the quality of training data. In particular, [12] presents a method using a trained distributed support vector machine. The approach in [13] exploits the relationships between statistical and geometric properties of the training data to learn the pattern of attacks. In [14], the online deep learning technique is employed to extract the features of attacks from the historical measurements and detect the attack in real-time.

Note that most of these existing state estimation methods are based on the assumption that the whole system is ideally balanced and synchronized (e.g., matched constant frequency) under normal operation, which is not always the case in real applications [15]. Moreover, existing works do not consider the cost of local samplings and data communications with the controller associated with attack detection.

### B. Contributions

We develop a joint state estimation and attack detection framework for a given bus in the smart grid, utilizing the sequential Monte Carlo (SMC) method to track the non-linear state-space model of a bus with frequency drift, and the CUSUM change detector for attack detection based on the state estimation. Exploiting historical observations, we further propose an adaptive sampling strategy such that the controller maintains a large sampling interval when the system is believed to be normal and quickly reduces the interval when the attack is considered likely. The adaptive sampling strategy efficiently reduces the rate of taking local measurements, while the detection delay performance is almost unaffected. Extensive simulation results demonstrate that even with the adaptive sampling strategy, our proposed method can quickly detect various attacks in a smart grid with high efficiency and adaptivity.

The remainder of the paper is organized as follows. Section II introduces the system models under both normal and attacked conditions, and presents the problem formulation. In Section III, we present the proposed state estimation and attack detection algorithm, as well as the adaptive sampling strategy. In Section IV, the proposed method is applied in a simulated power grid system and its performance is compared with the extended-Kalman-filter-based estimation method in [15] and conventional detection rules in [10]. Section V concludes the paper.

## II. System Model and Problem Statement

We begin by presenting the state-space model under the normal operation, and then specify the measurement models for several typical attacks in a real operating grid followed by the problem statement.

We consider a discrete-time state-space model for a power system. Let $\boldsymbol{x}(k)$ and $\boldsymbol{y}(k)$ denote the state and measurement vectors at the $k^{th}$ sampling instant, respectively. Then we have the following general state-space model:

$$\boldsymbol{x}(k+1) = f(\boldsymbol{x}(k), \Delta t(k)) + \boldsymbol{w}(k), \tag{1}$$

$$\boldsymbol{y}(k) = h(\boldsymbol{x}(k), \Delta t(k)) + \boldsymbol{v}(k), \tag{2}$$

where $f(\cdot)$ is the state transition function, $h(\cdot)$ is the measurement function, $\Delta t(k)$ is the time interval between the $k^{th}$ and the $(k+1)^{th}$ samples, $\boldsymbol{w}(k)$ is the process noise vector, and $\boldsymbol{v}(k)$ is the measurement noise vector. In the following subsections, detailed models under both normal and faulty conditions are presented.

### A. Model under Normal Conditions

For power system monitoring, sensors or meters are deployed to take measurements at various locations over time. Typically, these meters can send the measurement data (e.g., bus voltage) to a central controller via wired or wireless communication. In particular, we derive the state-space model based on the power grid voltage signal. The three-phase voltage signal at a given bus can be described by [16],

$$V_a(t) = V_a \cos(t\omega + \phi_a) + e_a(t), \tag{3}$$

$$V_b(t) = V_b \cos(t\omega + \phi_b) + e_b(t), \tag{4}$$

$$V_c(t) = V_c \cos(t\omega + \phi_c) + e_c(t), \tag{5}$$

where $t$ denotes the continuous time; $V_i$ and $\phi_i$ denote the voltage amplitude and the initial phase angle of signal $i \in \{a, b, c\}$; $\omega = 2\pi f_0$ where $f_0$ is typically equal to 50Hz or 60Hz. Note that, in practice, power systems are subject to frequency variation, e.g., $f_0$ varies from 60Hz to 61Hz. The additive noise $e_i(t)$ is modeled by a white zero-mean Gaussian process. Note that in an ideally balanced power system, $V_a = V_b = V_c$, and $\phi_b = \phi_a - \frac{2}{3}\pi$, $\phi_c = \phi_a + \frac{2}{3}\pi$.

We transform the noise-free signals in (3)-(5) into the $\alpha$-$\beta$ reference frame using Clarke transformation [16], i.e.,

$$[V_\alpha(t), V_\beta(t)]^T \triangleq \mathbf{T} \left[V_a(t), V_b(t), V_c(t)\right]^T, \tag{6}$$

where the Clarke transformation

$$\mathbf{T} \triangleq \frac{2}{3}\begin{bmatrix} 1 & -\frac{1}{2} & -\frac{1}{2} \\ 0 & \frac{\sqrt{3}}{2} & -\frac{\sqrt{3}}{2} \end{bmatrix}. \tag{7}$$

Then the noise-free signal components in the $\alpha$-$\beta$ frame are given by

$$\begin{aligned} V'_\alpha(t) &= \frac{2V_a}{3}\left[\cos(\omega t + \phi_a) - \frac{1}{2}\cos\left(\omega t + \phi_a - \frac{2}{3}\pi\right) - \frac{1}{2}\cos\left(\omega t + \phi_a + \frac{2}{3}\pi\right)\right] \\ &\triangleq V_\alpha \cos(\omega t + \phi_\alpha), \end{aligned} \tag{8}$$

$$\begin{aligned} V'_\beta(t) &= \frac{\sqrt{3}V_a}{3}\left[\cos\left(\omega t + \phi_a - \frac{2}{3}\pi\right) - \cos\left(\omega t + \phi_a + \frac{2}{3}\pi\right)\right] \\ &\triangleq V_\beta \cos(\omega t + \phi_\beta). \end{aligned} \tag{9}$$

Considering the noise terms denoted by $e_\alpha(t)$ and $e_\beta(t)$ in the $\alpha$-$\beta$ frame, the noisy signals after the transformation can be written as [16],

$$V_\alpha(t) = V_\alpha \cos(t\omega + \phi_\alpha) + e_\alpha(t), \tag{10}$$

$$V_\beta(t) = V_\beta \cos(t\omega + \phi_\beta) + e_\beta(t). \tag{11}$$

We define the following five state variables including the in-phase and quadrature signals along with the grid frequency based on (10) and (11) [17] [15]:

$$\begin{aligned} x_1(t) &= V_\alpha \cos(t\omega + \phi_\alpha), \\ x_2(t) &= V_\alpha \sin(t\omega + \phi_\alpha), \\ x_3(t) &= V_\beta \cos(t\omega + \phi_\beta), \\ x_4(t) &= V_\beta \sin(t\omega + \phi_\beta), \\ x_5(t) &= \omega. \end{aligned} \tag{12}$$

Recalling the general model in (1)-(2) and defining the state vector $\boldsymbol{x}(k) = [x_1(k), x_2(k), ..., x_5(k)]^T$ corresponding to the $k^{th}$ sampling instant, then we can write the following state transition equation [18]:

$$\begin{aligned} \boldsymbol{x}(k+1) &= f(\boldsymbol{x}(k), \Delta t(k)) + \boldsymbol{w}(k) \\ &= \begin{bmatrix} x_1(k)\cos\big(x_5(k)\cdot\Delta t(k)\big) - x_2(k)\sin\big(x_5(k)\cdot\Delta t(k)\big) \\ x_1(k)\sin\big(x_5(k)\cdot\Delta t(k)\big) + x_2(k)\cos\big(x_5(k)\cdot\Delta t(k)\big) \\ x_3(k)\cos\big(x_5(k)\cdot\Delta t(k)\big) - x_4(k)\sin\big(x_5(k)\cdot\Delta t(k)\big) \\ x_3(k)\cos\big(x_5(k)\cdot\Delta t(k)\big) + x_4(k)\sin\big(x_5(k)\cdot\Delta t(k)\big) \\ (1-\varepsilon)x_5(k) \end{bmatrix} + \boldsymbol{w}(k). \end{aligned} \tag{13}$$

It is assumed that the process noise $\boldsymbol{w}(k)$ under both normal and faulty conditions is white Gaussian with covariance matrix $\boldsymbol{W}$, i.e., $\boldsymbol{w}(k) \sim \mathcal{N}(\boldsymbol{0}, \boldsymbol{W})$. The parameter $\varepsilon$ characterizes the slowly time-varying characteristic of the grid frequency.

Define $\boldsymbol{y}(k) = [V_\alpha(k), V_\beta(k)]^T$. The measurement equation in (2) can then be written as

$$\begin{aligned} \boldsymbol{y}(k) &= h_0(\boldsymbol{x}(k)) + \boldsymbol{v}(k) \\ &\triangleq \mathbf{H}\boldsymbol{x}(k) + \boldsymbol{v}(k), \end{aligned} \tag{14}$$

where the measurement noise $\boldsymbol{v}(k) \sim \mathcal{N}(\boldsymbol{0}, \boldsymbol{R})$, and is uncorrelated with $\boldsymbol{w}(k)$, and

$$\mathbf{H} \triangleq \begin{bmatrix} 1 & 0 & 0 & 0 & 0 \\ 0 & 0 & 1 & 0 & 0 \end{bmatrix}. \tag{15}$$

### *B. Models for Typical Attacks*

In this paper, three typical attacks in a power grid are considered. Under each attack, the measurement equation varies while the state transition equation remains the same as (13).

*1) Denial of Service Attack:* A denial-of-service (DoS) attack is a cyber-attack which occurs when some components or resources become unavailable due to external adversaries. The DoS attack can be triggered by flooding the system with superfluous requests, which jams the communication channel and prevents legitimate data from being transmitted. In a smart grid, the DoS attack is modeled as the lack of measurement data at the central controller [19]. Thus, the observed signal at the controller is characterized by a Gaussian noise $\boldsymbol{u}(k)$ with mean $\boldsymbol{0}$ and covariance $\boldsymbol{U}$, i.e.,

$$\begin{aligned} \boldsymbol{y}(k) &= h_1(\boldsymbol{x}(k)) + \boldsymbol{v}(k) \\ &\triangleq \boldsymbol{u}(k). \end{aligned} \tag{16}$$

Note that, in general, the values of $\boldsymbol{U}$ and $\boldsymbol{R}$ may not be the same.

*2) Random Attack:* The random attack is modeled by an additive term, $\boldsymbol{y}_a(k)$, which manipulates the original meter readings at the sampling instant $k$ as

$$\begin{aligned} \boldsymbol{y}(k) &= h_2(\boldsymbol{x}(k)) + \boldsymbol{v}(k) \\ &\triangleq \mathbf{H}\boldsymbol{x}(k) + \boldsymbol{y}_a(k) + \boldsymbol{v}(k). \end{aligned} \tag{17}$$

We consider it an attack event of interest when $\|\boldsymbol{y}_a(k)\|^2 \geq a_0$, where $a_0$ is a predetermined lower bound on the magnitude of an attack.

*3) False Data Injection Attack:* The false data injection attack is induced when the compromised meters forge the events that do not occur. Assume that the attacker knows the topology of a power system and can control a subset of meters from different buses that are affected by the attack for the $k^{th}$ measurement. Now consider the attack affecting a given bus. The measurement model under attack at the given bus is [10]

$$\begin{aligned} \boldsymbol{y}(k) &= h_3(\boldsymbol{x}(k)) + \boldsymbol{v}(k) \\ &\triangleq \mathbf{H}\boldsymbol{x}(k) + \mathbb{1}(k)\boldsymbol{y}_a(k) + \boldsymbol{v}(k), \end{aligned} \tag{18}$$

where $\mathbb{1}(k) = 1$ if the given bus is attacked at time $k$, and $\mathbb{1}(k) = 0$ otherwise, and $\boldsymbol{y}_a(k)$ is the malicious input from the attacker. Comparing (17) and (18), it is seen that the false data injection attack can be intermittent whereas the random attack is persistent.

### *C. Problem Statement*

To jointly estimate the state of the bus and detect an attack based on voltage measurements, the controller needs to address the following problems.

*1) State Estimation:* The state variable $\boldsymbol{x}(k)$ cannot be measured directly. We need to estimate $\boldsymbol{x}(k)$ based on all the measurements $\boldsymbol{Y}(k) \triangleq [\boldsymbol{y}(1), \boldsymbol{y}(2)..., \boldsymbol{y}(k-1), \boldsymbol{y}(k)]$ reported by the meter at that bus.

*2) Detection Rule:* Our goal is to detect any attack as quickly as possible. After obtaining the state estimate at each sampling instant, according to a specific detection rule, the controller needs to make a decision on whether to trigger the attack alarm immediately or take more measurements to update the state estimate.

*3) Sampling Interval Adaptation:* Since the traditional uniform sampling strategy may lead to extensive measurements and high communication load between the controller and local meters, we propose to reduce the number of measurements by adjusting the sampling interval $\Delta t(k)$ adaptively. The basic idea is to maintain a large sampling interval during the normal operation and reduce the sampling interval upon observing irregular fluctuations.

## III. SMC-based State Estimation and Attack Detection

The proposed framework for state estimation and attack detection is illustrated in Fig. 1. The voltage readings from a certain bus are sequentially fed to the controller that runs the proposed SMC filtering algorithm. In particular, the SMC filter iteratively estimates the distribution of the state vector. The estimated state is then fed to a CUSUM change detector to detect the attack. If no attack alarm is triggered, the controller adjusts the sampling interval and waits for the next measurement.

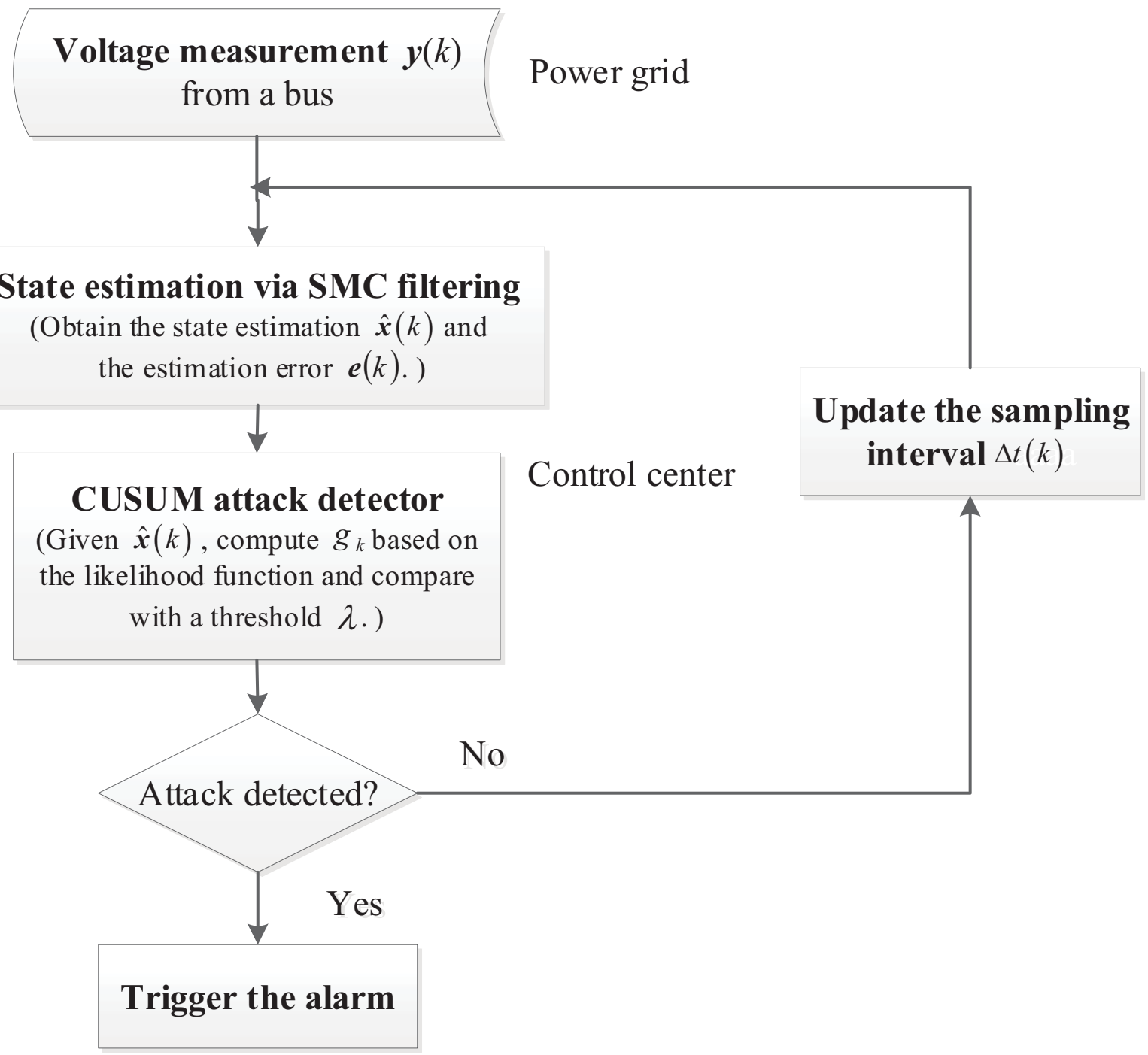

Fig. 1. Flow chart of the proposed online detector of attacks on bus meters based on voltage measurement.

In the following subsections, we first present the SMC-based algorithm to estimate the system state given the sampling interval $\Delta t(k)$, and then specify the CUSUM-based attack detection rule. Finally, we present a strategy to adaptively adjust $\Delta t(k)$ to reduce the measurement cost and the communication load between the meter and the controller.

### A. State Estimation via SMC Algorithm

Our proposed attack detection approach is based on the state estimation by the SMC method. In SMC [20], a set of weighted samples are used to approximate an underlying distribution that is to be estimated. And the samples and their associated weights are sequentially updated based on the new measurements.

The proposed SMC-based state estimator is summarized in Algorithm 1. The algorithmic details are presented as follows.

- **Initialization**

**Algorithm 1** SMC-based State Estimator

```
Initialization: k = 0, draw N samples, {x^(j)(0)}_{j=1}^N, according to the prior distribution N(x^0, Σ^0). Set the weight w̄_j(0) = 1/N for all j.
for k = 1, 2, ... do
    Generate the current samples x^(j)(k) according to (21).
    Update the weight w_j(k) according to (22) and normalize it.
    Compute the state and error estimates, x̂(k) and e(k), according to (24) and (25).
    Perform resampling if N̂_eff is below a given threshold.
end for
```

For $k=0$, draw $N$ initial samples, $\left\{\boldsymbol{x}^{(j)}(0)\right\}_{j=1}^{N}$, from the prior probability density function characterized by $\boldsymbol{x}^0$ and $\boldsymbol{\Sigma}^0$, i.e.,

$$\boldsymbol{x}^{(j)}(0) \sim \mathcal{N}(\boldsymbol{x}^0, \boldsymbol{\Sigma}^0), \quad j = 1, 2, ..., N. \tag{19}$$

Set initial weights $w_j(0) = 1/N$ for all $j$.

- **Online State Estimation**

During the online phase, the controller sequentially updates the state estimate. An SMC update step consists of *sample generation*, *weight update*, and *resampling*, as highlighted in Algorithm 1.

*1) Sample generation:* The basic idea of SMC is to perform the sequential importance sampling (SIS). At each time, $N$ samples $\left\{\boldsymbol{x}^{(j)}(0)\right\}_{j=1}^{N}$ are drawn from some trial distribution $\pi\left(\boldsymbol{x}^{(j)}(k)|\boldsymbol{X}^{(j)}(k-1), \boldsymbol{Y}(k)\right)$ with $\boldsymbol{X}^{(j)}(k-1) \triangleq \left\{\boldsymbol{x}^{(j)}(1), \boldsymbol{x}^{(j)}(2), ..., \boldsymbol{x}^{(j)}(k-1)\right\}$. Here we choose the state transition density as the trial distribution, i.e.,

$$\pi\left(\boldsymbol{x}^{(j)}(k)|\boldsymbol{X}^{(j)}(k-1), \boldsymbol{Y}(k)\right) \triangleq p\big(\boldsymbol{x}^{(j)}(k)|\boldsymbol{x}^{(j)}(k-1)\big). \tag{20}$$

Hence according to (13),

$$\boldsymbol{x}^{(j)}(k) \sim \mathcal{N}\left(f(\boldsymbol{x}^{(j)}(k-1), \Delta t(k-1)), \boldsymbol{W}\right). \tag{21}$$

*2) Weight update:* The corresponding weight $w_j(k)$ for sample $\boldsymbol{x}^{(j)}(k)$ is calculated by

$$\begin{aligned} w_j(k) &\propto \overline{w}_j(k-1)\frac{p(\boldsymbol{y}(k)|\boldsymbol{x}^{(j)}(k))p(\boldsymbol{x}^{(j)}(k)|\boldsymbol{x}^{(j)}(k-1))}{\pi(\boldsymbol{x}^{(j)}(k)|\boldsymbol{X}^{(j)}(k-1), \boldsymbol{Y}(k))} \\ &\propto \overline{w}_j(k-1)p\left(\boldsymbol{y}(k)|\boldsymbol{x}^{(j)}(k)\right) \\ &\propto \overline{w}_j(k-1)\cdot\exp\left[-\frac{1}{2}\left(\boldsymbol{y}(k)-\mathbf{H}\boldsymbol{x}^{(j)}(k)\right)^T\boldsymbol{R}^{-1}\left(\boldsymbol{y}(k)-\mathbf{H}\boldsymbol{x}^{(j)}(k)\right)\right], \end{aligned} \tag{22}$$

where the normalized weight $\overline{w}_j(k)$ is given as

$$\overline{w}_j(k) = \frac{w_j(k)}{\sum_{j=1}^{N} w_j(k)}. \tag{23}$$

Given the current weighted samples $\{x^{(j)}(k), \overline{w}_j(k)\}_{j=1}^{N}$, we can estimate the state vector and the error vector as

$$\widehat{\boldsymbol{x}}(k) = \sum_{j=1}^{N} \overline{w}_j(k) \boldsymbol{x}^{(j)}(k), \tag{24}$$

$$\boldsymbol{e}(k) = \sum_{j=1}^{N} \overline{w}_j(k) \left[\boldsymbol{y}(k) - \mathbf{H}\boldsymbol{x}^{(j)}(k)\right]. \tag{25}$$

*3) Resampling:* The resampling step aims to avoid the problem of degeneracy of the SMC algorithm, that is, the situation that all but one of the importance weights are close to zero [21] [22]. The basic solution is to retain the samples with high weights and discard the samples with low weights.

The resampling is implemented only when the effective number of particles $N_{\text{eff}}$ is below a predetermined threshold $N_{\text{thr}}$. An estimate of $N_{\text{eff}}$ is given by

$$\widehat{N}_{\text{eff}} = \frac{1}{\sum_{i=1}^{N} \left(\overline{w}_j(k)\right)^2}, \tag{26}$$

which reflects the variation of the weights [21]. If $\widehat{N}_{\text{eff}}$ is less than a given threshold, $N_{\text{thr}}$, we perform resampling to obtain $N$ new samples, $\left\{\widetilde{\boldsymbol{x}}^{(j)}(k)\right\}_{j=1}^{N}$, with equal weights $\widetilde{w}_j(k) = 1/N$. That is, draw $N$ samples from the current sample set with probabilities proportional to the corresponding weights.

### *B. Attack Detection via CUSUM Test*

As given by (14)-(18), the statistical models for measurements before and after the attack occurrence are the basis to formulate a change detection problem which aims at the quickest reaction to the sudden change [23] [24]. The basic strategy is to utilize the sequential measurements to achieve high time resolution, and thus minimize the detection delay subject to the constraint on the false alarm period. At the $k^{th}$ sampling instant, the detector takes the current measurement $\boldsymbol{y}(k)$, and computes a decision statistic $g_k$, based on which it decides whether or not to declare an attack.

To facilitate the description of the proposed detection rule, we first write the conditional probability density functions of measurements corresponding to the null hypothesis $\theta_0$ and the alternative hypothesis $\theta_a \in \{\theta_1, \theta_2, \theta_3\}$ ($\theta_1$ for DoS attack, $\theta_2$ for random attack, and $\theta_3$ for false data injection attack) as

$$p_{\theta_0}\left(\boldsymbol{y}(k)\left|\boldsymbol{x}(k)\right.\right) = \frac{1}{\sqrt{(2\pi)^2\left(\det \boldsymbol{R}_0\right)}} \exp\big(-\frac{1}{2}\boldsymbol{\varepsilon}_{k,\boldsymbol{\theta}_0}^T {\boldsymbol{R}_0}^{-1} \boldsymbol{\varepsilon}_{k,\theta_0}\big), \tag{27}$$

$$p_{\theta_1}\left(\boldsymbol{y}(k)\left|\boldsymbol{x}(k)\right.\right) = \frac{1}{\sqrt{(2\pi)^2\left(\det \boldsymbol{R}_1\right)}} \exp\big(-\frac{1}{2}\boldsymbol{\varepsilon}_{k,\theta_1}^T \boldsymbol{R}_1^{-1} \boldsymbol{\varepsilon}_{k,\theta_1}\big), \qquad \theta_a = \theta_1 \tag{28}$$

$$p_{\theta_2}\left(\boldsymbol{y}(k)\left|\boldsymbol{x}(k), \boldsymbol{y}_a(k)\right.\right) = \frac{1}{\sqrt{(2\pi)^2\left(\det \boldsymbol{R}_2\right)}} \exp\big(-\frac{1}{2}\boldsymbol{\varepsilon}_{k,\theta_2}^T \boldsymbol{R}_2^{-1} \boldsymbol{\varepsilon}_{k,\theta_2}\big), \qquad \theta_a = \theta_2 \tag{29}$$

$$p_{\theta_3}\left(\boldsymbol{y}(k)\left|\boldsymbol{x}(k), \boldsymbol{y}_a(k), \mathbb{1}(k)\right.\right) = \frac{1}{\sqrt{(2\pi)^2\left(\det \boldsymbol{R}_3\right)}} \exp\big(-\frac{1}{2}\boldsymbol{\varepsilon}_{k,\theta_3}^T \boldsymbol{R}_3^{-1} \boldsymbol{\varepsilon}_{k,\theta_3}\big), \qquad \theta_a = \theta_3 \tag{30}$$

where

$$\boldsymbol{\varepsilon}_{k,\theta_0} \triangleq \boldsymbol{y}(k) - \mathbf{H}\boldsymbol{x}(k), \tag{31}$$

$$\boldsymbol{\varepsilon}_{k,\theta_1} \triangleq \boldsymbol{y}(k), \tag{32}$$

$$\boldsymbol{\varepsilon}_{k,\theta_2} \triangleq \boldsymbol{y}(k) - \mathbf{H}\boldsymbol{x}(k) - \boldsymbol{y}_a(k), \tag{33}$$

$$\boldsymbol{\varepsilon}_{k,\theta_3} \triangleq \boldsymbol{y}(k) - \mathbf{H}\boldsymbol{x}(k) - \mathbb{1}(k)\boldsymbol{y}_a(k), \tag{34}$$

and $\{\boldsymbol{R}_0, ..., \boldsymbol{R}_3\}$ denote the corresponding covariances. Given the models in (14)-(18), $\boldsymbol{R}_0 = \boldsymbol{R}_2 = \boldsymbol{R}_3 = \boldsymbol{R}$ and $\boldsymbol{R}_1 = \boldsymbol{V}$.

For each attack type $\theta_a \in \{\theta_1, \theta_2, \theta_3\}$, the occurrence of the attack can be detected at time $T$ via the following sequential change detection procedure, called CUSUM test [23],

$$g_k = (g_{k-1} + l_k)^+, \tag{35}$$

$$T = \inf\left\{k : g_k \geq \lambda\right\}, \tag{36}$$

where

$$l_k = \frac{1}{2}\ln\frac{\det\left(\boldsymbol{R}_0\right)}{\det\left(\boldsymbol{R}_a\right)} + \frac{1}{2}\boldsymbol{\varepsilon}_{k,\theta_0}^T {\boldsymbol{R}_\mathbf{0}}^{-1}\boldsymbol{\varepsilon}_{k,\theta_0} - \frac{1}{2}\boldsymbol{\varepsilon}_{k,\theta_a}^T \boldsymbol{R}_a^{-1}\boldsymbol{\varepsilon}_{k,\theta_a}, \tag{37}$$

with $\boldsymbol{R}_a \in \{\boldsymbol{R}_1, \boldsymbol{R}_2, \boldsymbol{R}_3\}$ denoting the covariance under different alternative hypotheses, and $\lambda$ is a threshold.

Given a predetermined false alarm period $\gamma$, the threshold is approximated by $\lambda \approx \ln \gamma$ [23]. In our case of attack detection, the stopping time $T$ is the first time that $g_k$ exceeds the threshold, indicating the occurrence of an attack and terminating the current detection cycle.

Since the true state $\boldsymbol{x}(k)$, the exact value of $\boldsymbol{y}_a(k)$ or $\mathbb{1}(k)$ are unknown, we need to replace the terms $\varepsilon_{k,\theta_0}$ and $\varepsilon_{k,\theta_a}$ in (37) with their estimates based on the output of the SMC state estimator, i.e.,

$$\widehat{\varepsilon}_{i,\theta_0} = \boldsymbol{y}(k) - \mathbf{H}\widehat{\boldsymbol{x}}(k), \tag{38}$$

$$\widehat{\varepsilon}_{i,\theta_1} = \boldsymbol{y}(k), \qquad \theta_a = \theta_1, \tag{39}$$

$$\widehat{\varepsilon}_{i,\theta_2} = \boldsymbol{y}(k) - \mathbf{H}\widehat{\boldsymbol{x}}(k) - \widehat{\boldsymbol{y}}_a(k), \qquad \theta_a = \theta_2, \tag{40}$$

$$\widehat{\varepsilon}_{i,\theta_3} = \boldsymbol{y}(k) - \mathbf{H}\widehat{\boldsymbol{x}}(k) - \widehat{\mathbb{1}(k)\boldsymbol{y}_a(k)}, \qquad \theta_a = \theta_3, \tag{41}$$

where $\widehat{\boldsymbol{x}}(k)$ is given in (1).

For the random attack model in (40), the attack sequence $\boldsymbol{y}_a(k)$ is estimated as

$$\widehat{\boldsymbol{y}}_a(k) = \begin{cases} \boldsymbol{e}(k), & \|\boldsymbol{e}(k)\|^2 \geq a_0 \\ \frac{\sqrt{2}}{2}\left[a_0, a_0\right]^T & \|\boldsymbol{e}(k)\|^2 < a_0 \end{cases} \tag{42}$$

with $\boldsymbol{e}(k)$ given in (25). (42) implies that the attack sequence is approximated by the estimation error $\boldsymbol{e}(k)$ as long as $\|\boldsymbol{e}(k)\|^2$ exceeds the lower bound of the attack magnitude of interest.

For the false attack model in (41), the attack term $\mathbb{1}(k)\boldsymbol{y}_a(k)$ is estimated as

$$\widehat{\mathbb{1}(k)\boldsymbol{y}_a(k)} = \begin{cases} \boldsymbol{e}(k), & \text{if } \|\boldsymbol{e}(k)\|^2 \geq a_0 \text{ and } -\frac{1}{2}\widehat{\varepsilon}_{i,\theta_3}^T \boldsymbol{R}_3^{-1}\widehat{\varepsilon}_{i,\theta_3}\Big|_{\widehat{\mathbb{1}(k)\boldsymbol{y}_a(k)}=\boldsymbol{e}(k)} > -\frac{1}{2}\widehat{\varepsilon}_{i,\theta_0}^T \boldsymbol{R}_3^{-1}\widehat{\varepsilon}_{i,\theta_0} \\ \frac{\sqrt{2}}{2}\left[a_0, a_0\right]^T & \text{if } \|\boldsymbol{e}(k)\|^2 < a_0 \text{ and } -\frac{1}{2}\widehat{\varepsilon}_{i,\theta_3}^T \boldsymbol{R}_3^{-1}\widehat{\varepsilon}_{i,\theta_3}\Big|_{\widehat{\mathbb{1}(k)\widehat{\boldsymbol{y}}_a(k)}=\frac{\sqrt{2}}{2}[a_0,a_0]^T} > -\frac{1}{2}\widehat{\varepsilon}_{i,\theta_0}^T \boldsymbol{R}_3^{-1}\widehat{\varepsilon}_{i,\theta_0} \\ 0, & \text{otherwise.} \end{cases} \tag{43}$$

The estimator in (43) ensures that, given the $k^{th}$ sample, the bus is estimated as normal or attacked such that $\widehat{l_k}$ is maximized.

### C. Adaptive Sampling Strategy

Conventionally, the sampling interval $\Delta t(k)$ is a constant, i.e., $\Delta t(k) = \Delta$ for all $k$. Here we propose an adaptive sampling strategy that adjusts the next sampling interval based on past observations. Intuitively, the sampling interval should be relatively large when the system is under normal operation and it should quickly decrease when an attack is perceived as likely. To that end, we adopt the idea from the congestion control for the network transmission control protocol (TCP), i.e., the additive increase/multiplicative decrease scheme for obtaining a proper data package size [25]. Specifically, the proposed sampling method has two phases described as follows.

*1) Normal Operation Phase (Initialization):* During the normal operation, we aim to quickly find a default sampling interval which balances between the resource cost and the measurement resolution. Suppose that the minimum achievable sampling interval is $T_m$. We start with the sampling interval $\Delta t(i=0) = T_m$, and $\Delta t(i)$ is doubled after each measurement $i$ as long as the error does not exceed a predetermined threshold, $d_0$, i.e.,

$$\Delta t(i+1) = 2 \times \Delta t(i), \text{ if } \|\boldsymbol{e}(i)\|^2 \leq d_0, \ i = 0, 1, 2, ... \tag{44}$$

where $\boldsymbol{e}(i)$ is given by (25). Denote the maximum sampling interval satisfying (44) as $\Delta t_0$, which is set as the default sampling interval under normal operation.

*2) Online Attack Detection Phase:* When the controller believes that the estimation error is reasonable for normal operations, the sampling interval can be increased as

$$\Delta t(k+1) = \min\left\{\lfloor a \times \Delta t_0 + \Delta t(k) \rceil, b \times \Delta t_0\right\}, \text{ if } \|\boldsymbol{e}(k)\|^2 \leq d_1, \ k = 0, 1, 2, ... \tag{45}$$

where the function $\lfloor x \rceil$ rounds $x$ to the nearest multiple of $T_m$, the coefficient $a < 1$ determines the growth rate of the sampling interval, $b > 1$ is a predetermined integer defining the upper bound of the interval to ensure the detection performance, and the threshold $d_1 > d_0$.

When the controller observes the potential of an attack occurrence, it should increase the sampling frequency significantly to achieve higher measurement resolution until an attack alarm is triggered, i.e.,

$$\Delta t(k+1) = \max\left\{\lfloor c \times \Delta t(k) \rceil, T_m\right\}, \text{ if } \|\boldsymbol{e}(k)\|^2 > d_2, \ k = 0, 1, 2, ... \tag{46}$$

where $d_2 > d_1$ is the threshold, the coefficient $c < 1$ controls the decreasing rate and the sampling interval cannot be further refined when $T_m$ is reached.

After detecting the attack, the sampling interval is reset to the default value $\Delta t_0$, and adjusted by (45) and (46) in the next cycle.

Fig. 2 illustrates an example of the proposed adaptive sampling method, where the coefficients,

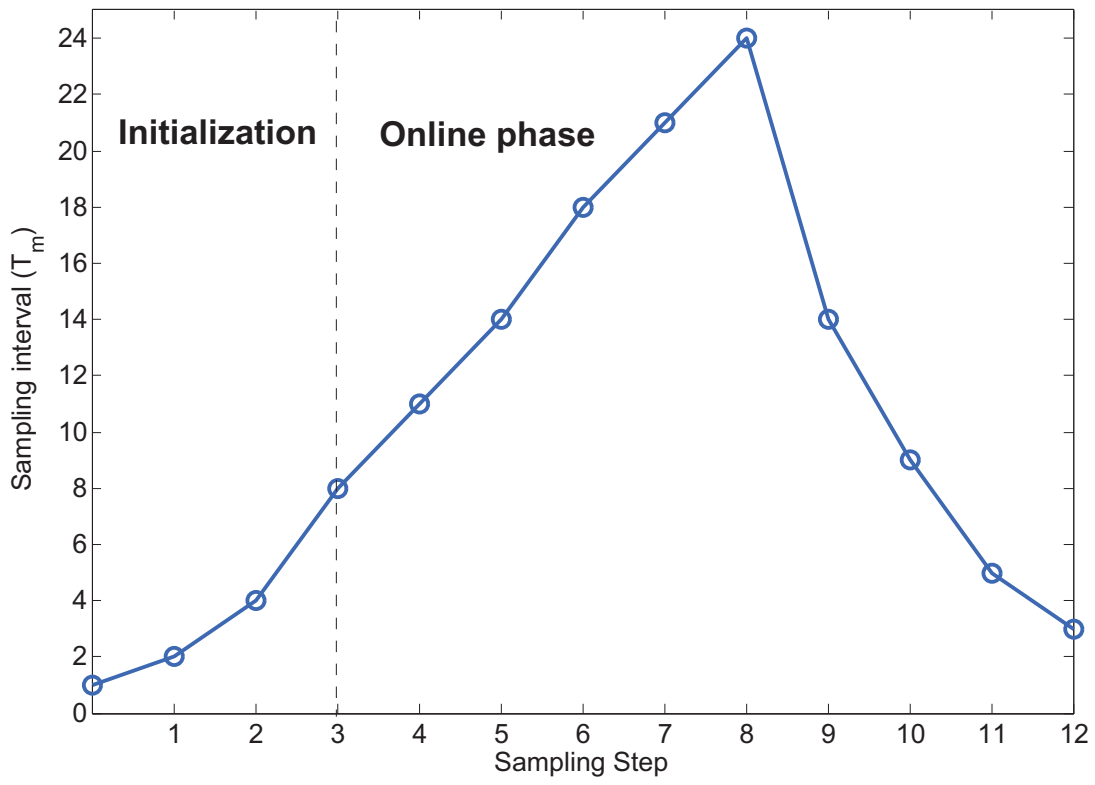

Fig. 2. Illustration of the adaptive sampling strategy.

$a = 0.4$, $b = 3$ and $c = 0.6$, with three thresholds, $d_0 = 1.5 \times 10^{-5}$, $d_1 = 2 \times 10^{-5}$ and $d_2 = 3 \times 10^{-5}$. In the initialization phase, the default sampling interval $\Delta t_0 = 8T_m$ is quickly reached in the third step. In the online phase, the sampling interval rises from $8T_m$ to $3\Delta t_0 = 24T_m$. When a large deviation is observed in the $8^{th}$ step, the sampling interval is adjusted to $3T_m$ after the $12^{th}$ sample.

Finally, the complete procedure of our proposed attack detection method is summarized in Algorithm 3.

## IV. Simulation Results

In this section, we present the performance of the proposed method in detecting typical attacks at a given bus in a power system, i.e., denial of service attacks, random attacks and false data injection attacks.

For each type of attacks, we first present the estimation performance of SMC, and then examine the delay performance of the proposed attack detector with the adaptive sampling scheme. To demonstrate the effectiveness of the proposed algorithm, we also compare its delay performance with that of the extended-Kalman-filter-based estimation method [15] in combination with conventional attack detectors [10]. Finally, we evaluate the amount of resource saved by the proposed adaptive sampling strategy. Note that, for a fair comparison, the average delay given a certain false alarm period is measured by the equivalent number of samples between the

**Algorithm 2** Proposed Attack Detector

1: **Initialization**:
2: Obtain the default sampling interval $\Delta t_0$ given $d_0$ according to (44).
3: For SMC, obtain the initial samples $\{\boldsymbol{x}^{(j)}(0)\}_{j=1}^{N}$, and set the threshold for resampling.
4: For the CUSUM test, set the lower bound for an attack of interest, $a_0$, and the attack decision thresholds, $\lambda$.
5: For the sampling interval adaptation, set the coefficients, $a$, $b$ and $c$, and the thresholds, $d_1$ and $d_2$.
6: **Online attack detection**:
7: **for** $k = 1, 2, ...$ **do**,
8: 1) Take the $k^{th}$ measurement.
9: 2) Run one step of SMC update in Algorithm 1.
10: 3) Compute $g_k$ as in (35).
11: **if** $g_k \geq \lambda$ **then**
12: Trigger the attack alarm and break the loop.
13: **else**
14: Update the next sampling interval $\Delta t(k)$ according to (45) and (46).
15: **end if**
16: Wait $\Delta t(k)$.
17: $k = k + 1$.
18: Loop back to Line 8.
19: **end for**

attack occurrence and the attack alert using the default sampling frequency $\Delta t_0$ [1].

### A. Simulation Setup

In our experiments, we simulate different types of faults in a power system using Matlab. The detailed setups to implement the proposed method and other methods for comparison are described below.

*1) Proposed Method:* A sinusoidal voltage signal with random Gaussian noise is generated. For the system model, the amplitudes of the three-phase voltage signal $V_a = V_b = V_c = 1$, and the initial phase angles, $\phi_a$, $\phi_b$ and $\phi_c$, are set as $0$, $-\frac{2\pi}{3}$ and $\frac{2\pi}{3}$, respectively. The grid frequency $f_0$ linearly varies from 60Hz to 60.5Hz to examine the adaptation to frequency variations. The covariance matrix of the process noise $\boldsymbol{W} = \begin{bmatrix} \sigma^2 \mathbf{I}_{4\times 4} & \mathbf{0} \\ \mathbf{0} & 10^{-7} \end{bmatrix}$, where $\sigma^2$ is set to $10^{-3}$ or $10^{-5}$ for comparison, and $\mathbf{I}_{4\times 4}$ is the $4 \times 4$ identity matrix. The parameter $\varepsilon$ reflecting the slow varying of the grid frequency is set to $10^{-16}$. The measurement noise covariances, $\boldsymbol{R} = 10^{-3}\mathbf{I}_{2\times 2}$ and $\boldsymbol{U} = 1.5 \times 10^{-3}\mathbf{I}_{2\times 2}$. The nominal fundamental frequency $f_0 = 60$Hz, and the sampling frequency of local meters $f_m = 1/T_m = 10^6$Hz.

[1]Suppose the time interval between the attack occurrence and the alarm is $T_{delay}$, the delay is measured by $N_{sample} = T_{delay}/\Delta t_0$.

For the SMC algorithm, the number of particles before and after resampling are set as $N = 500$. The initial samples $\{\boldsymbol{x}^{(j)}(0)\}_{j=1}^{N}$ are draw from The prior probability density function $\mathcal{N}(\boldsymbol{x}^0, \boldsymbol{\Sigma}^0)$ where $\boldsymbol{x}^0 = [0,0,0,0,60]^T$, and $\boldsymbol{\Sigma}^0 = \begin{bmatrix} \mathbf{I}_{4\times 4} & \mathbf{0} \\ \mathbf{0} & 10^{-7} \end{bmatrix}$. The lower bound for classifying an attack of interest is set as $a_0 = 1.5 \times \sigma$ with different process noise covariances. For the adaptive sampling strategy, the coefficients for sampling interval adjustment are set as $a = 0.4$, $b = 3$, and $c = 0.6$. During the initialization, the threshold $d_0 = 1.5 \times \sigma$ in (44). For the online interval adjustment, $d_1 = 2 \times \sigma$ and $d_2 = 3 \times \sigma$. Under the uniform sampling method, the sampling interval is fixed to $\Delta t_0$.

*2) Methods for Comparison:* To tackle the nonlinear estimation problem, the methods in [15] utilizes the extended Kalman filter (EKF). We compare the estimation result with that of the EKF to demonstrate the higher accuracy of our SMC method.

To evaluate the performance of attack detection, we compare our CUSUM-based detector with conventional ones [10], namely, the $\chi^2$ detector and Euclidean distance detector, under uniform sampling with the sampling interval fixed to $\Delta t_0$. Moreover, the performance of another detector based on the well-known likelihood ratio test with adaptive sampling is also compared. Note that the following detection methods do not need prior knowledge about the attack patterns, and can trigger an alarm without differentiating the attack type.

$\chi^2$ **Detector**: The $\chi^2$ detector computes the chi-square distributed statistic $\chi^2(k) \triangleq \boldsymbol{e}(k)^T \boldsymbol{S}(k)^{-1}\boldsymbol{e}(k)$, where the residual covariance matrix $\boldsymbol{S}(k)$ is computed as

$$\boldsymbol{S}(k) = \sum_{j=1}^{N} \left[\mathbf{H}\boldsymbol{x}^{(j)}(k) - \widehat{\boldsymbol{y}}(k)\right] \left[\mathbf{H}\boldsymbol{x}^{(j)}(k) - \widehat{\boldsymbol{y}}(k)\right]^T. \tag{47}$$

with

$$\widehat{\boldsymbol{y}}(k) = \sum_{j=1}^{N} \mathbf{H}\boldsymbol{x}^{(j)}(k). \tag{48}$$

Then a $\chi^2$ test [10] will declare the occurrence of an attack after the $k_0^{th}$ sample with

$$k_0 = \min\left\{k : \chi^2(k) \geq \lambda_1\right\}, \tag{49}$$

where $\lambda_1 = \chi^2_{d=1}(\eta)$ is the detection threshold with $1-\eta$ denoting the target level of confidence and the degrees of freedom $d = 2 - 1 = 1$.

**Euclidean-Distance (E-D) Detector**: Since the false data injection attack can be carefully crafted to pass conventional statistical detection, e.g., $\chi^2$ detector, another detection rule based on the Euclidean distance metric is proposed in [10]. Here we adopt a windowed Euclidean distance measure among the past $W$ true voltage measurements, $\{V_i(k-W+1), ..., V_i(k)\}$, and the corresponding estimated values, $\left\{\widetilde{V}_i(k-W+1), ..., \widetilde{V}_i(k)\right\}$, of phase $i \in \{a, b, c\}$, reconstructed from the state estimates, i.e.,

$$d(V_i, \widetilde{V}_i, k) = \sqrt{\sum_{k'=k-W+1}^{k} \left(V_i(k') - \widetilde{V}_i(k')\right)}. \tag{50}$$

An attack alarm is triggered after the $k_0^{th}$ sample with

$$k_0 = \min\left\{k : d(V_i, \widetilde{V}_i, k) \geq \lambda_2\right\}, \forall i \in \{a, b, c\}. \tag{51}$$

**Log-likelihood-ratio (LLR) Detector**: This windowed detection rule takes the latest $W$ likelihood-ratio evaluations into account. Given a threshold $\lambda_3$, the decision of an attack is made after the $k_0^{th}$ sample where

$$k_0 = \min\left\{k : \sum_{k'=k-W+1}^{k} \widehat{l}_{k'} \geq \lambda_3\right\}, \tag{52}$$

For a fair comparison of the detection performance, the thresholds, $\lambda$ (for our method), $\lambda_1$, $\lambda_2$ and $\lambda_3$ (for the conventional detectors in [10], and the LLR detector), are tuned to satisfy the same target false alarm periods. The window sizes for both the E-D detector and the LLR detector are set as $W = 10$.

### *B. Detection of DoS Attack*

To simulate the DoS attack, the controller is prevented from observing the upcoming local measurements after the attack which occurs at $t = 0.05$s and lasts for the following 0.05s. To illustrate the estimation process, the estimated sinusoid output in phase $a$, $\widehat{V}_a(t)$ (reconstructed from the state variables), is presented in Fig. 3, in comparison with the true input voltage signal given the large measurement noise ($\sigma^2 = 10^{-3}$). Before the attack occurrence at $t = 0.05$s, the estimated values by the SMC algorithm closely track the true signal. After $t = 0.05$s, the gap between the estimated signal and the true signals becomes obvious. It can be observed that the estimate by the EKF generally results in larger errors in comparison with our SMC method.

We evaluate the average attack detection delay performance of different methods as a function of the false alarm period $\gamma$ (in terms of the equivalent number of samples under default sampling frequency). The delay was the average of 5000 simulations. Fig. 4 (a) and (b) shows the average delay performances for detecting the DoS attack with different process noises. On one hand, given a specific false alarm period, the proposed algorithm with uniform sampling intervals has the shortest average delays in both cases, implying higher efficiency of our method. On the other hand, when the sampling interval $\Delta(k)$ is adaptively adjusted, the delay performance is only affected slightly. Even given a large false alarm period which is the desired case for real applications, the difference between the two sampling strategies is still very small. Compared with the EKF-based $\chi^2$ detector with uniform sampling and the LLR-test detector, the proposed approach has less average delays given the same false alarm periods even with adaptive sampling.

Fig. 5 illustrates the adaptive sampling process by showing the average length of the sampling interval during different time periods (false alarm period $\gamma = 10^3$). For each time period specified in the legend, the average sampling interval $\overline{T}_x$ is obtained by computing the ratio between the length of the specified time period $T_x$, and the total number of samples at the controller, $n_x$, during the corresponding period, i.e., $\overline{T}_x = T_x/n_x$. Before the attack occurs, the controller starts with the default sampling frequency (after initialization), and the sampling interval under normal operation is around $20T_m$. After the attack occurrence, the controller quickly shrinks the sampling interval to increase the measurement resolution. For the last 5 samples before the detection decision is made, the sampling interval gets closer to the smallest sampling period corresponding to the highest achievable sampling frequency of the local meter ($\Delta t(k) = 2.3T_m \sim 2.5T_m$). By comparing results under different measurement noises, we can conclude that the average sampling interval is larger in general when the noise level is lower.

As shown in Table I, the amount of resource saving by the adaptive sampling is evaluated based on the average sampling interval, $\overline{T}_a = \frac{1}{2}\left(\overline{T}_{normal} + \overline{T}_{attack}\right)$, in comparison with the uniform sampling approach. Explicitly, the resource saving percentage is given by $\left(1 - T_m/\overline{T}_a\right) \times 100\%$. For different process noise levels ($\sigma^2 = 10^{-3}$ and $10^{-5}$), the percentages of resource saving are all above 92.3% with different false alarm periods. In general, the adaptive sampling method can save more resources when the measurement noise is low, which agrees with our observation in Fig. 5.

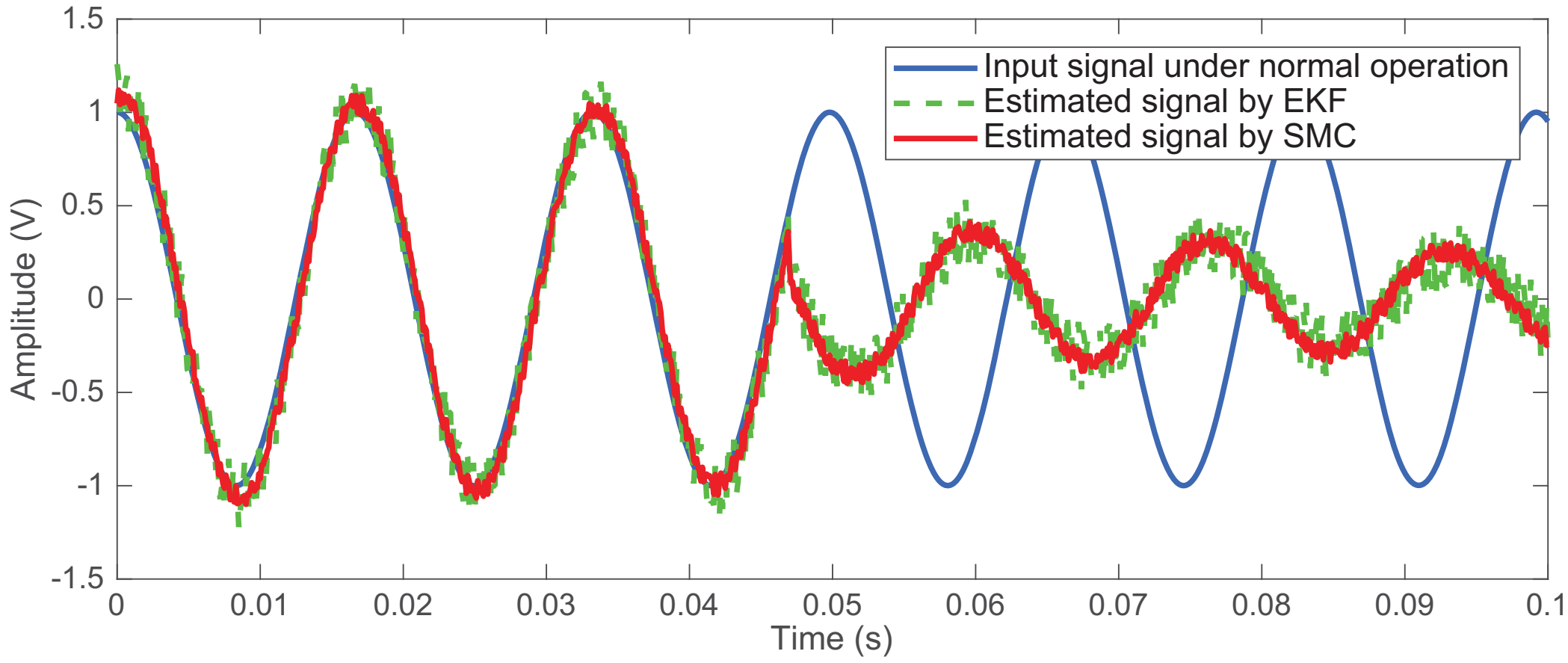

Fig. 3. Estimates of $V_a(t)$ by SMC and EKF under the DoS attack ($\sigma^2 = 10^{-3}$).

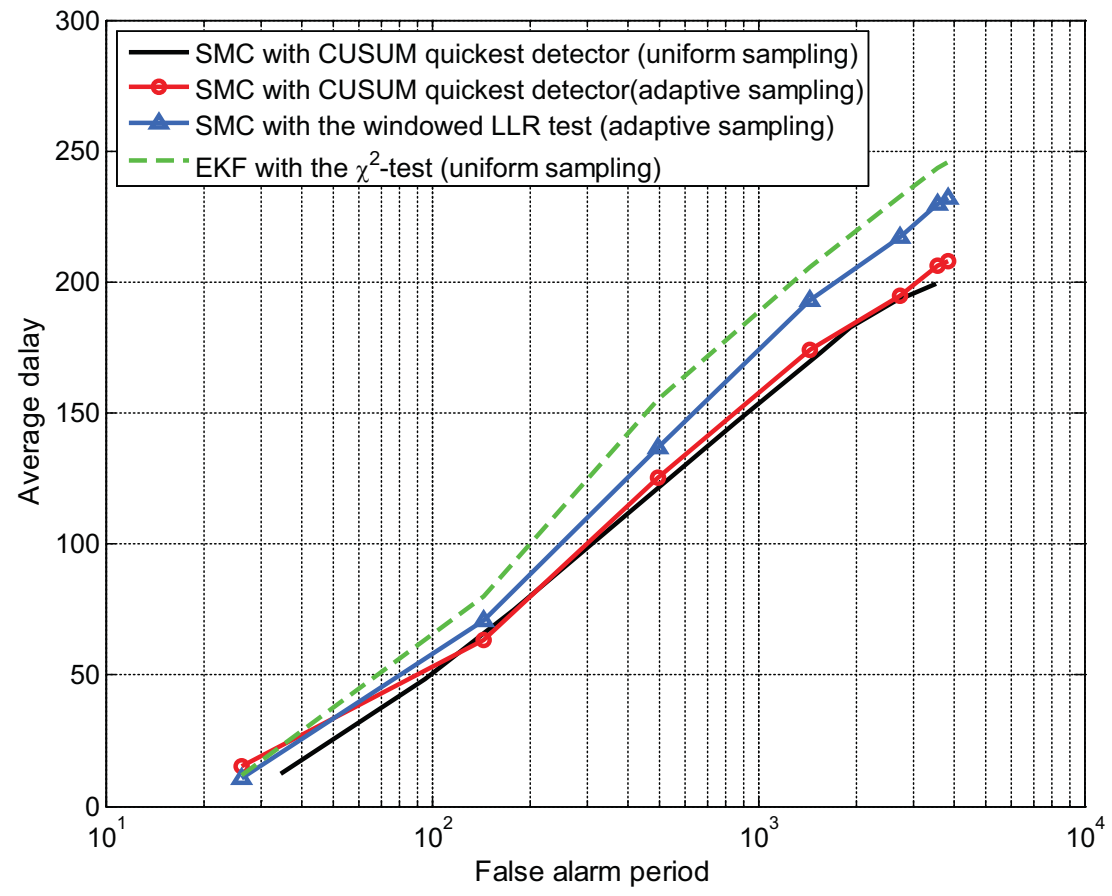

(a) Detection delay for DoS attacks ($\sigma^2 = 10^{-5}$).

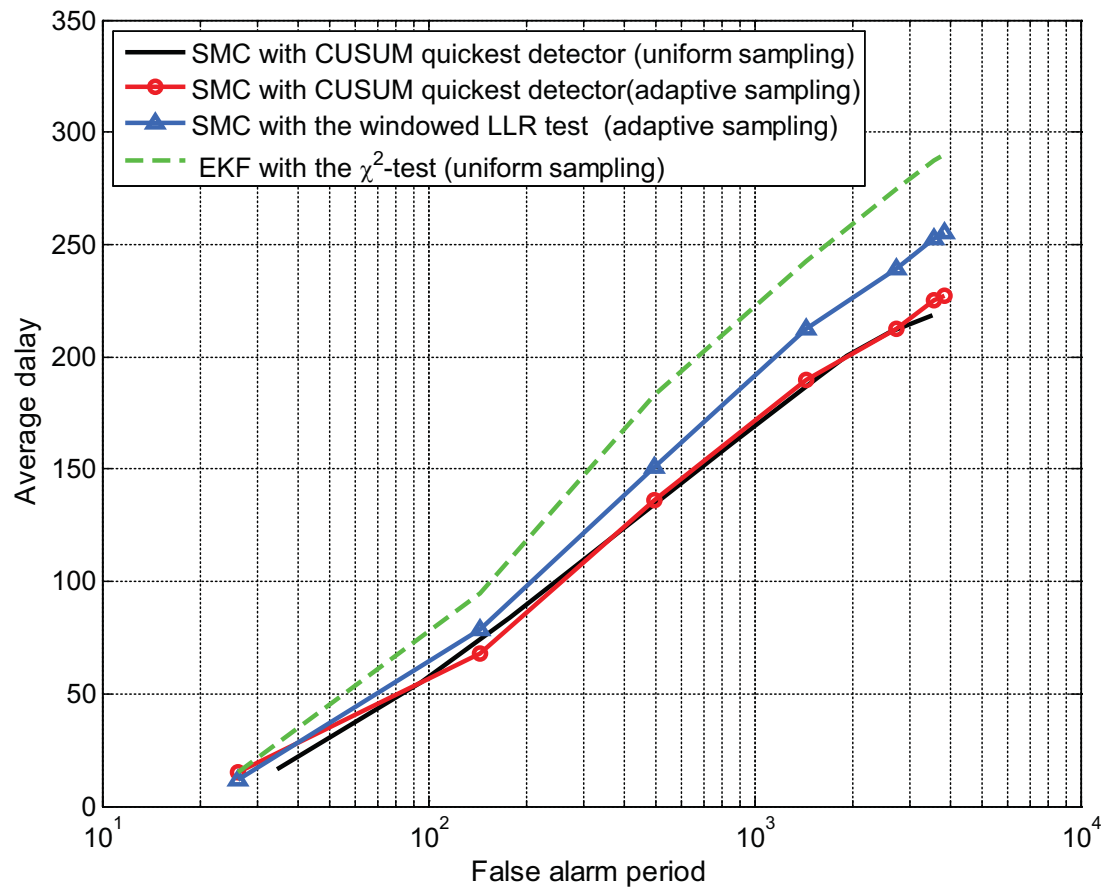

(b) Detection delay for DoS attacks ($\sigma^2 = 10^{-3}$).

Fig. 4. Comparison of detection delay performances under DoS attacks.

TABLE I
PERCENTAGE OF RESOURCE SAVING BY ADAPTIVE SAMPLING UNDER DOS ATTACKS.

| $\gamma$ | Resource saving | |
|---|---|---|
| | $\sigma^2 = 10^{-5}$ | $\sigma^2 = 10^{-3}$ |
| $10^2$ | 93.6% | 92.6% |
| $5 \times 10^2$ | 93.3% | 92.4% |
| $10^3$ | 93.1% | 92.3% |

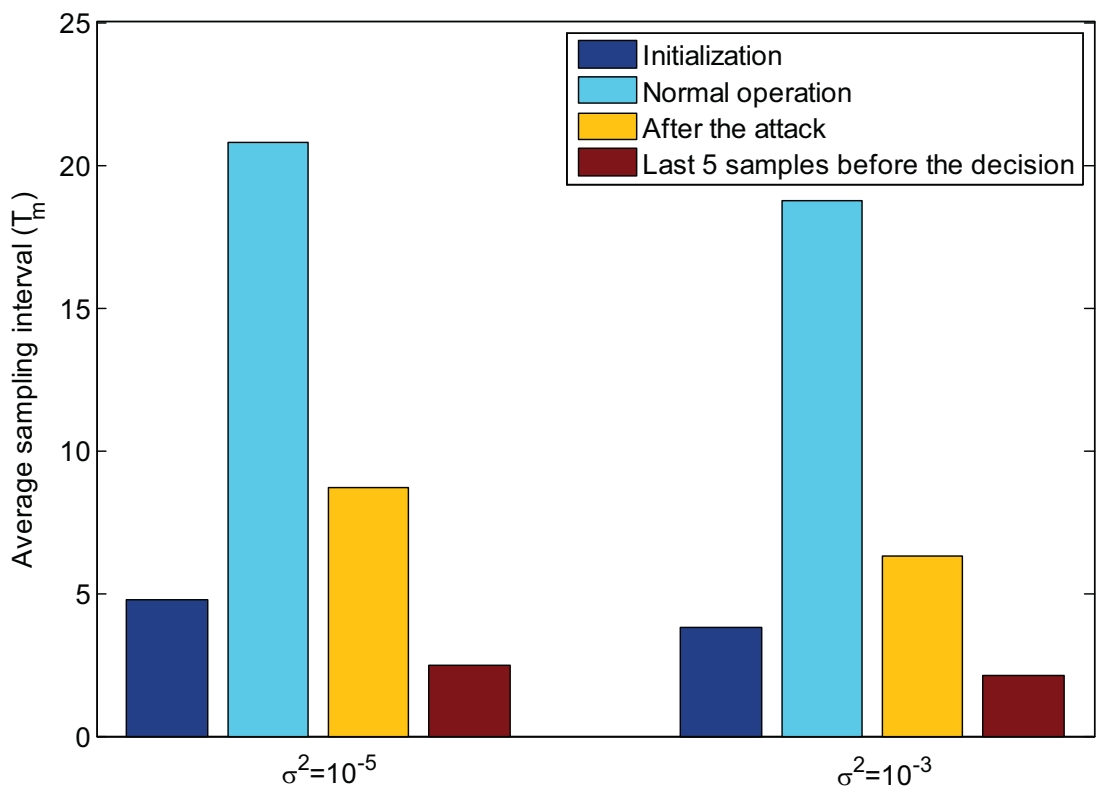

Fig. 5. Average sampling intervals for detecting DoS attacks under different noise conditions.

## *C. Detection of Random Attack*

Since the the random attack occurs when the controller obtains manipulated measurement data rather than the true local observations, in our simulation, the faulty signals can be generated by adding an attack vector $\boldsymbol{y}_a(k)$ to the original observations. The attack sequence for $V_a(t)$, $V_b(t)$ and $V_c(t)$ are sequentially generated by $1.5\cos((2\pi \times 65)t_k - \pi/4)$, where $t_k$ denotes the time instant for the $k^{th}$ measurement, and is transformed into $\alpha - \beta$ space accordingly to obtain the attack sequence $\boldsymbol{y}_a(k)$. Note that all the observations from $V_a(t)$, $V_b(t)$ and $V_c(t)$ are affected by the same attack sequence simultaneously.

Fig. 6 shows the estimated $V_a(t)$, where the continuous attack takes effect after $t = 0.05$s, i.e., all the following observations at controller are distorted by the attack. Similar to Fig. 3, the attack results in an abrupt gap between the estimated and the true signals, and our SMC-based method leads to smaller errors.

The average delay performance for detecting the random attack is shown in Fig. 7 (a) and (b) under different levels of measurement noise. Again, the proposed adaptive sampling performs similarly to uniform sampling, and exhibits less detection delays in comparison with the EKF-based method using $\chi^2$ detector. Compared with the result for detecting the DoS attack, the delay becomes larger in general, which coincides with our observation from Fig. 6 where the differences between the estimated signal under attack and the normal input signal are generally smaller than that in Fig. 3.

Fig. 8 shows the average length of the sampling interval during the corresponding time period. Similar to Fig. 5, the controller tends to maintain a large sampling interval during the normal

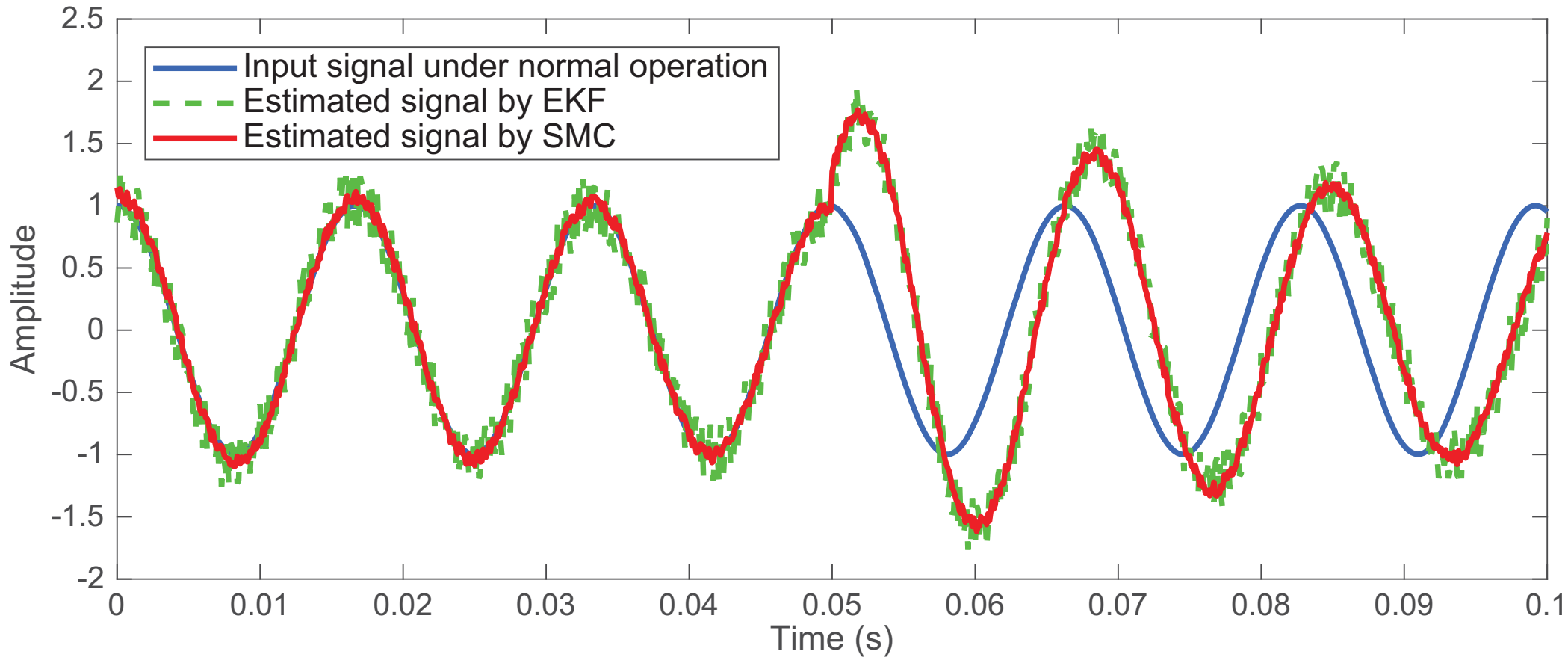

Fig. 6. Estimates of $V_a(t)$ by SMC and EKF under the random attack ($\sigma^2 = 10^{-3}$).

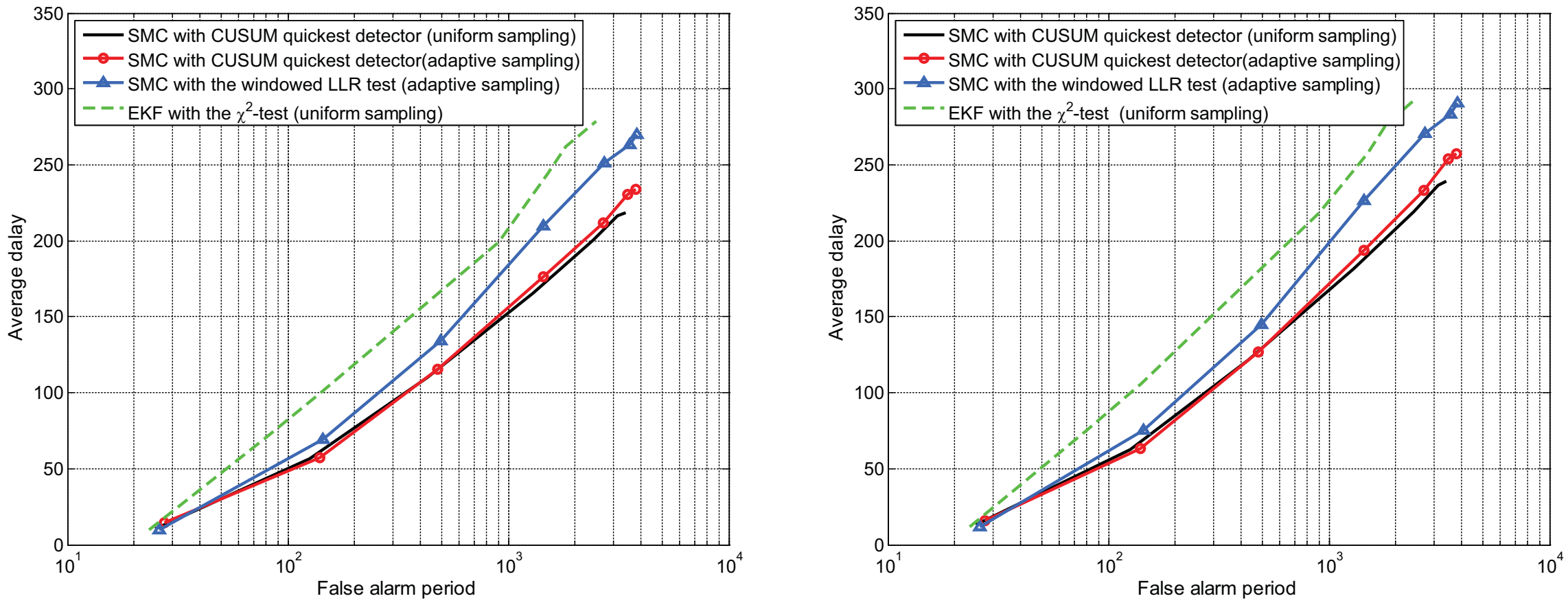

(a) Detection delay for random attacks $\sigma^2 = 10^{-5}$.

(b) Detection delay for random attacks $\sigma^2 = 10^{-3}$.

Fig. 7. Comparison of detection delay performances under random attacks.

operation and quickly reduces the interval after the attack occurrence. Table II demonstrates that a great amount of resource can be saved by implementing the adaptive sampling method (>92.1%).

### *D. Detection of False Data Injection Attack*

The simulated data injection sequence is generated by $\boldsymbol{y}_a(k+1) = \boldsymbol{y}_a(k) + \boldsymbol{z}$, where the random elements in $\boldsymbol{z}$ are independently generated from the uniform distribution in $[0.05, 0.15]$ and the initial attack vector is $[0, 0]^T$. Note that false data injection differs from the previous two types of attacks as the obtained measurements at the controller at different sampling instants can be either normal or faulty after the first attack occurrence, depending on whether the current state variable is affected by the attack. In our simulation, the value of the indicator function $\mathbb{1}(k)$ at different sampling instant are independently generated from the binomial distribution, such that $\mathbb{1}(k) = 1$ with probability of $p = 0.8$.

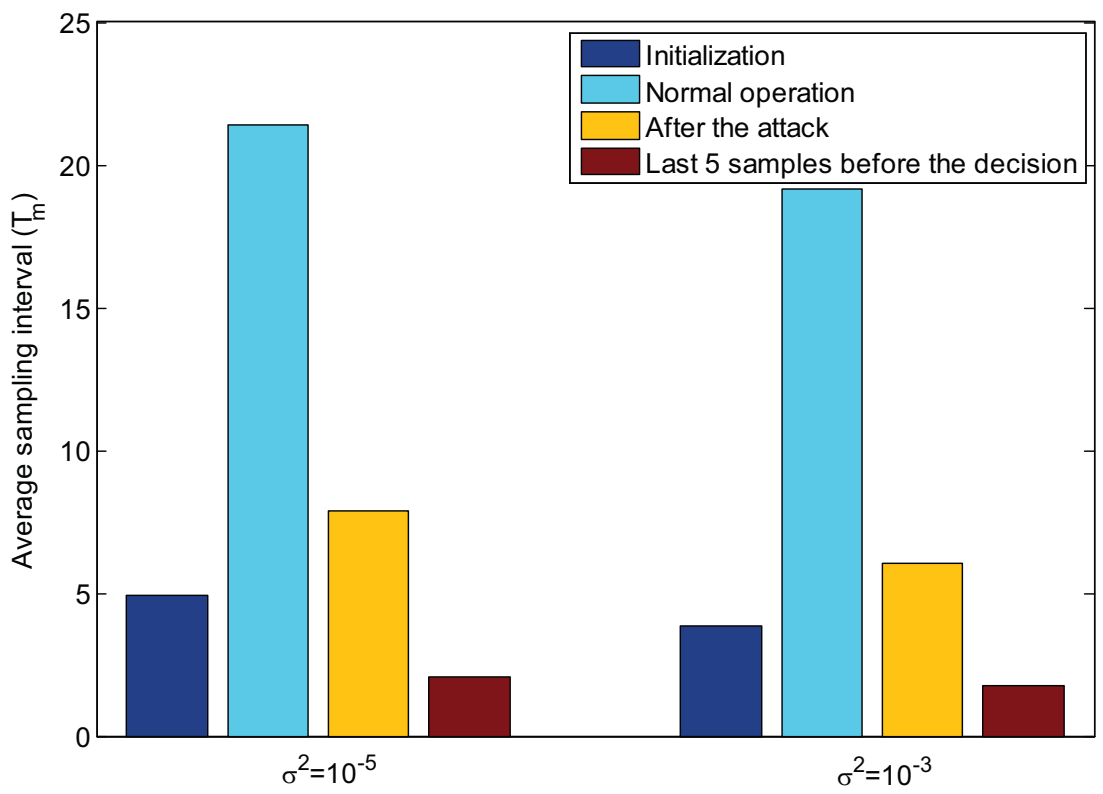

Fig. 8. Average sampling intervals for detecting random walk attacks under different noise conditions.

TABLE II
PERCENTAGE OF RESOURCE SAVING BY ADAPTIVE SAMPLING UNDER RANDOM ATTACKS.

| $\gamma$ | Resource saving | |
|---|---|---|
| | $\sigma^2 = 10^{-5}$ | $\sigma^2 = 10^{-3}$ |
| $10^2$ | 93.5% | 92.7% |
| $5 \times 10^2$ | 93.3% | 92.5% |
| $10^3$ | 93.0% | 92.1% |

The estimation result is shown in Fig. 9, where the estimate of $V_a(t)$ gradually deviates from the true signal after the attack occurrence at $t = 0.05$s. Compared with the previous two types of attacks, the gap between the estimates and the true signals under the false data injection attack does not become obvious until $t = 0.06$, implying a longer detection delay.

Fig. 10 (a) and (b) show the average delay performance for detecting the false data injection attack given different measurement noises. In response to the attack occurrence at $t = 0.05$s, the proposed method outperforms the others with less delays given the same false alarm period and the adaptive sampling scheme. Fig. 11 summaries the average sampling interval for detecting the false data injection attack, where we can observe that the average interval is smaller than

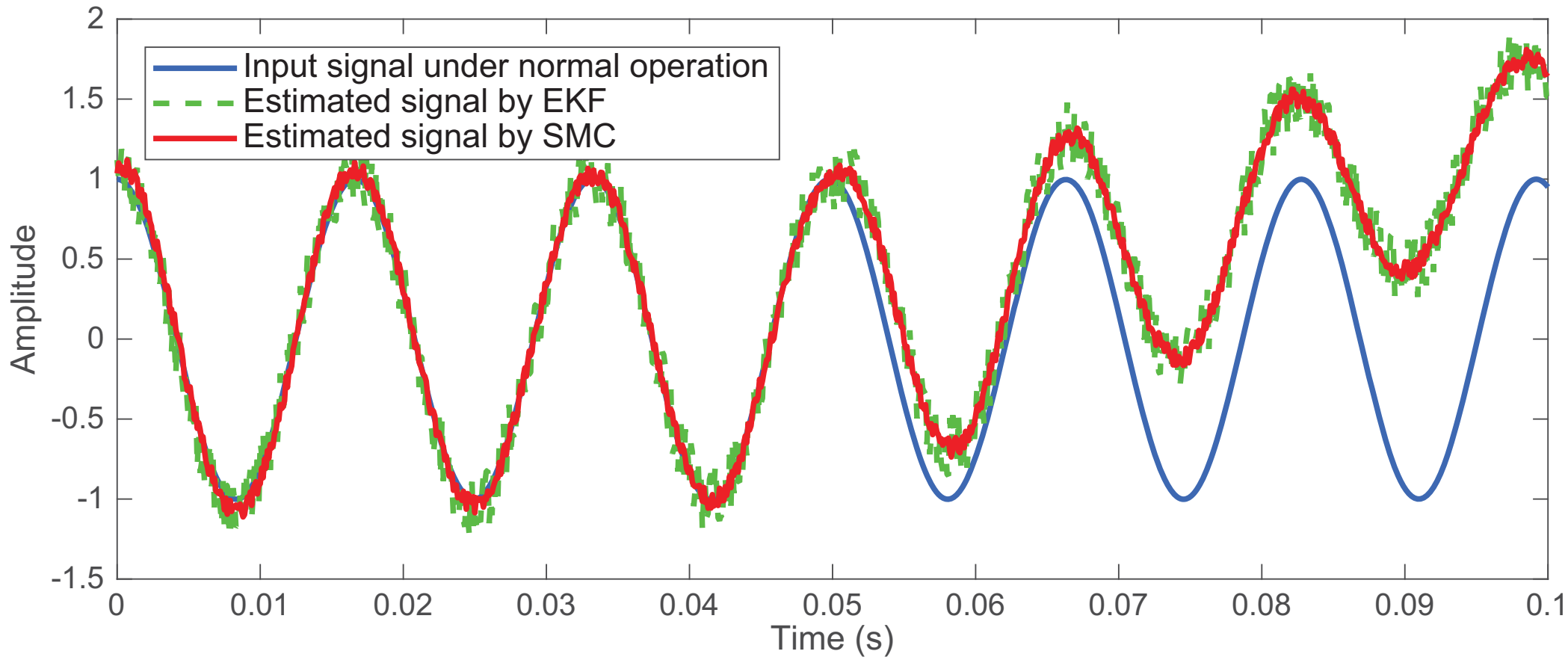

Fig. 9. Estimates of $V_a(t)$ by SMC and EKF under the false data injection attack ($\sigma^2 = 10^{-3}$).

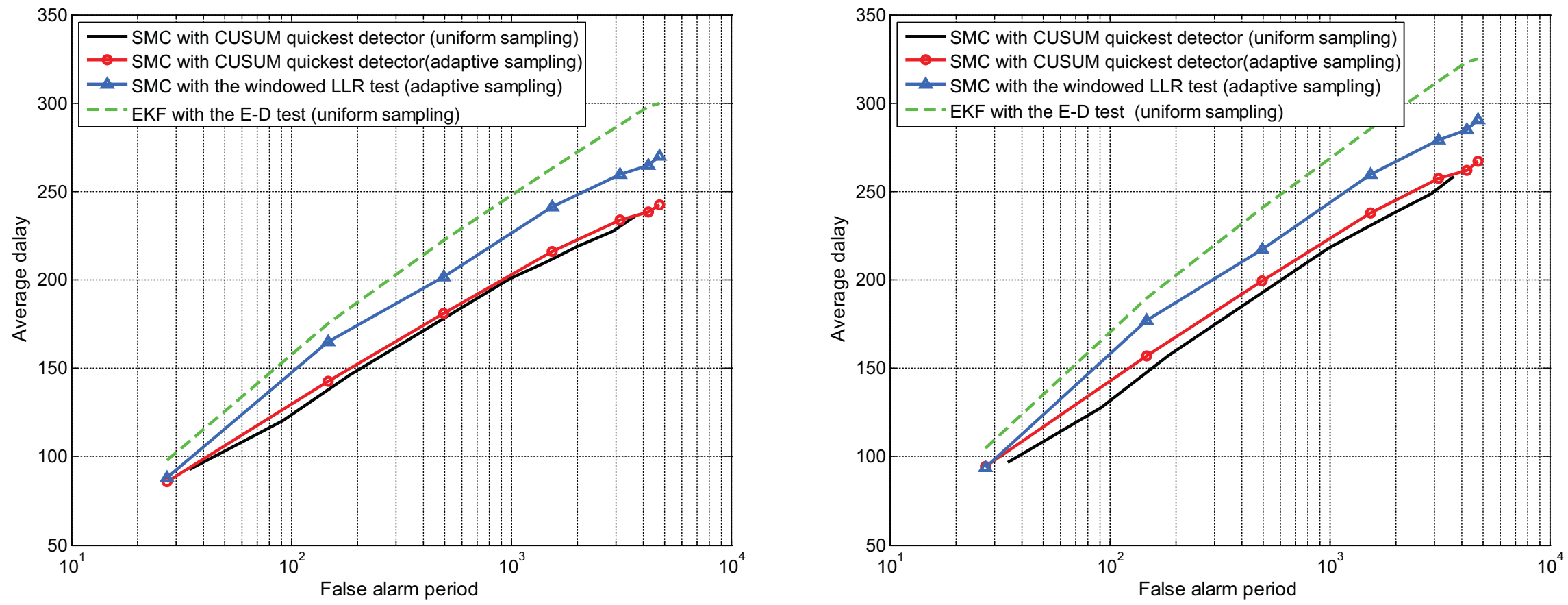

(a) Detection delay for false data injection attacks $\sigma^2 = 10^{-5}$.
(b) Detection delay for false data injection attacks $\sigma^2 = 10^{-3}$.

Fig. 10. Comparison of detection delay performances under false data injection attacks.

that for the other two types of attack in general. Table. III presents the result of the average resource saving, where the values varies from 92.0% to 92.9% under difference conditions.

## V. Conclusions

We have developed a joint SMC-based state estimation and CUSUM change detection method that quickly detects different attacks at a three-phase bus in a smart grid. The proposed SMC filtering tracks the non-linear state-space model when the grid frequency is slowing drifting, and the CUSUM-type detector can be adapted to various unknown attack patterns. Considering the resource cost on sampling and communication, an adaptive sampling strategy is proposed such that the controller adjusts the sampling interval based on the bus state estimation. The proposed method is implemented to detect three typical attacks in a power system, and compared with existing methods. The simulation results demonstrate that our method can quickly detect various faults without any prior knowledge. Furthermore, the proposed adaptive sampling strategy reduces the rate of taking measurements significantly while the detection delay is almost unaffected, which makes it suitable to be applied in real power grids.

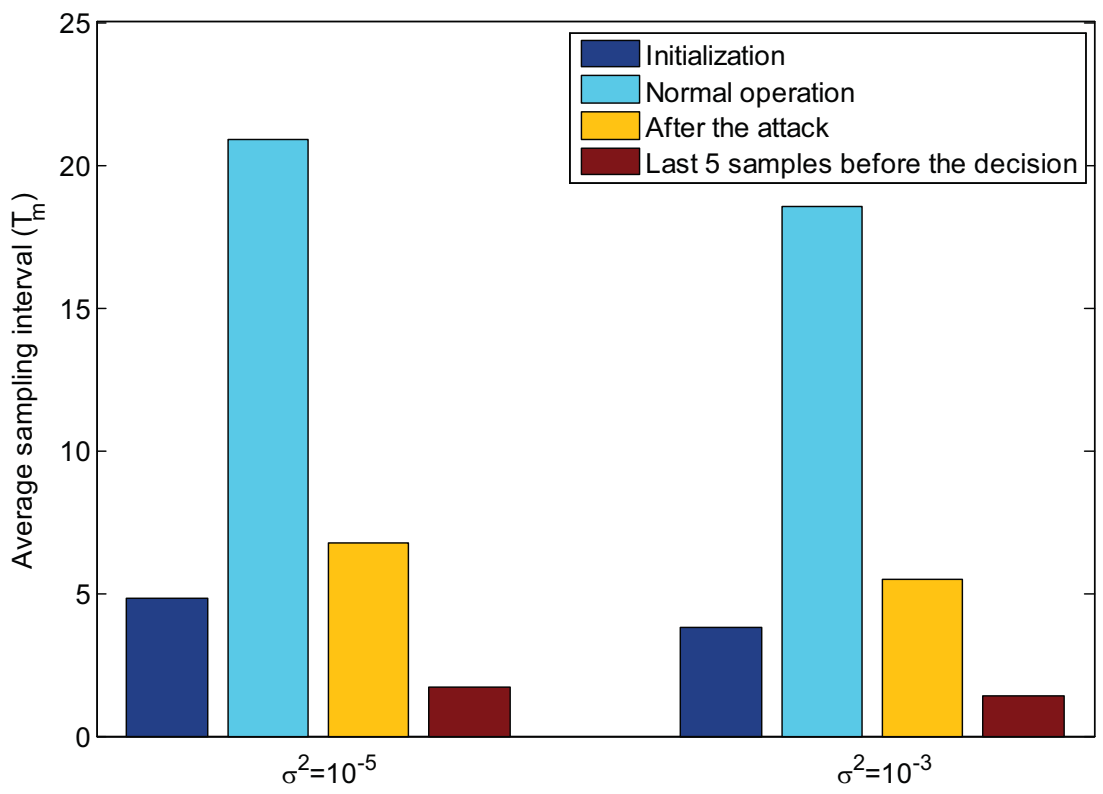

Fig. 11. Average sampling intervals for detecting false data injection attacks under different noise conditions.

TABLE III
PERCENTAGE OF RESOURCE SAVING BY ADAPTIVE SAMPLING UNDER FALSE DATA INJECTION ATTACKS.

| $\gamma$ | Resource saving | |
|---|---|---|
| | $\sigma^2 = 10^{-5}$ | $\sigma^2 = 10^{-3}$ |
| $10^2$ | 92.9% | 92.6% |
| $5 \times 10^2$ | 92.7% | 92.2% |
| $10^3$ | 92.6% | 92.0% |